# Haze Formation in Warm H$_2$-rich Exoplanet Atmospheres

Chao He[1*], Sarah M. Hörst[1,2], Nikole K. Lewis[3], Xinting Yu[1,4], Julianne I. Moses[5], Patricia McGuiggan[6], Mark S. Marley[7], Eliza M.-R. Kempton[8], Caroline V. Morley[9], Jeff A. Valenti[2], & Véronique Vuitton[10]

[1] Department of Earth and Planetary Sciences, Johns Hopkins University, Baltimore, MD, USA che13@jhu.edu

[2] Space Telescope Science Institute, Baltimore, MD, USA

[3] Department of Astronomy and Carl Sagan Institute, Cornell University, Ithaca, New York 14853, USA

[4] Department of Earth and Planetary Sciences, University of California Santa Cruz, CA 95064, USA

[5] Space Science Institute, Boulder, CO, USA

[6] Department of Materials Science and Engineering, Johns Hopkins University, Baltimore, MD, USA

[7] NASA Ames Research Center, Mountain View, CA, USA

[8] Department of Astronomy, University of Maryland, College Park, MD, USA

[9] Department of Astronomy, the University of Texas at Austin, Austin, TX, USA

[10] Institut de Planétologie et d'Astrophysique de Grenoble, Université Grenoble Alpes, CNRS, Grenoble, FR

**Abstract:**

New observing capabilities coming online over the next few years will provide opportunities for characterization of exoplanet atmospheres. However, clouds/hazes could be present in the atmospheres of many exoplanets, muting the amplitude of spectral features. We use laboratory simulations to explore photochemical haze formation in $H_2$-rich exoplanet atmospheres at 800 K with metallicity either 100 and 1000 times solar. We find that haze particles are produced in both simulated atmospheres with small particle size (20 to 140 nm) and relative low production rate ($2.4 \times 10^{-5}$ to $9.7 \times 10^{-5}$ mg cm$^{-3}$ h$^{-1}$), but the particle size and production rate is dependent on the initial gas mixtures and the energy sources used in the simulation experiments. The gas phase mass spectra show that complex chemical processes happen in these atmospheres and generate new gas products that can further react to form larger molecules and solid haze particles. Two $H_2$-rich atmospheres with similar C/O ratios (~0.5) yield different haze particles size, haze production rate, and gas products, suggesting both the elemental abundances and their bonding environments in an atmosphere can significantly affect the photochemistry. There is no methane ($CH_4$) in our initial gas mixtures, although $CH_4$ is often believed to be required to generate organic hazes. However, haze production rates from our experiments with different initial gas mixtures indicate that $CH_4$ is neither required to generate organic hazes nor necessary to promote the organic haze formation. The variety and relative yield of the gas products indicate that CO and $N_2$ enrich chemical reactions in $H_2$-rich atmospheres.

1. Introduction

A majority of exoplanets discovered in the last two decades have sizes between that of Earth and Neptune (i.e., super-Earths and mini-Neptunes). The Transiting Exoplanet Survey Satellite (TESS) mission, launched in 2018, is discovering many new exoplanets in this size range. Theoretical models have shown that these planets could have diverse atmospheric compositions (Elkins-Tanton & Seager 2008; Miller-Ricci et al. 2009; Schaefer et al. 2012; Moses et al. 2013a; Hu & Seager 2014; Ito et al. 2015; Venot et al. 2015). Future telescopes, including the James Webb Space Telescope (JWST, scheduled to launch in 2021) and ground-based extremely large telescopes (planned to operate in the late 2020s), will characterize these atmospheres. However, observations and laboratory simulations indicate that condensate clouds and/or photochemical hazes could be ubiquitous in the atmospheres of exoplanets, affecting their observed spectra (Knutson et al. 2014a, 2014b; Kreidberg et al. 2014; Dragomir et al. 2015; Sing et al. 2016; Lothringer et al. 2018; He et al. 2018a; Hörst et al. 2018a).

Aerosol particles (clouds and/or hazes) can impact both chemical and physical processes in planetary atmospheres and on planetary surfaces. Photochemically generated hazes are of particular interest as they may serve as a source of organic materials for potential chemical evolution of life on a planet (Khare et al. 1986, McDonald et al. 1994, Hörst et al. 2012, Gautier et al. 2014, Sebree et al. 2018). However, photochemical haze formation in exoplanet atmospheres remains poorly understood because the atmospheric composition and temperature regimes are relatively unexplored. The composition and observational impact of haze particles is also unclear.

Laboratory experimental simulations have advanced our understanding of haze formation in planetary atmospheres in the Solar System, e.g., Titan (Cable et al. 2012). Such simulations can also probe photochemistry in exoplanet atmospheres (He et al. 2018a, 2018b, 2019; Hörst et al. 2018a; Berry et al. 2019b; Fleury et al. 2019; Moran et al. 2020). Previously, we conducted a series of laboratory simulation experiments with atmospheric compositions relevant to super-Earths and mini-Neptunes at temperatures ranging from 300 K to 600 K. We reported production rates (He et al. 2018a; Hörst et al. 2018a) and size distributions (He et al. 2018a, 2018b) of solid haze particles that formed, as well as the gas and solid phase chemistry (He et al. 2019, Moran et al. 2020). Here we explore haze formation in warmer $H_2$-rich atmospheres.

In the current study, we extend our previous experimental matrix (He et al. 2018a, 2018b, 2019; Hörst et al. 2018a) to a temperature of 800 K because many known exoplanets occupy this temperature regime, including super-Earths and mini-Neptunes such as HD 97658 b (Van Grootel et al. 2004) and GJ 436 b (Butler et al. 2004).

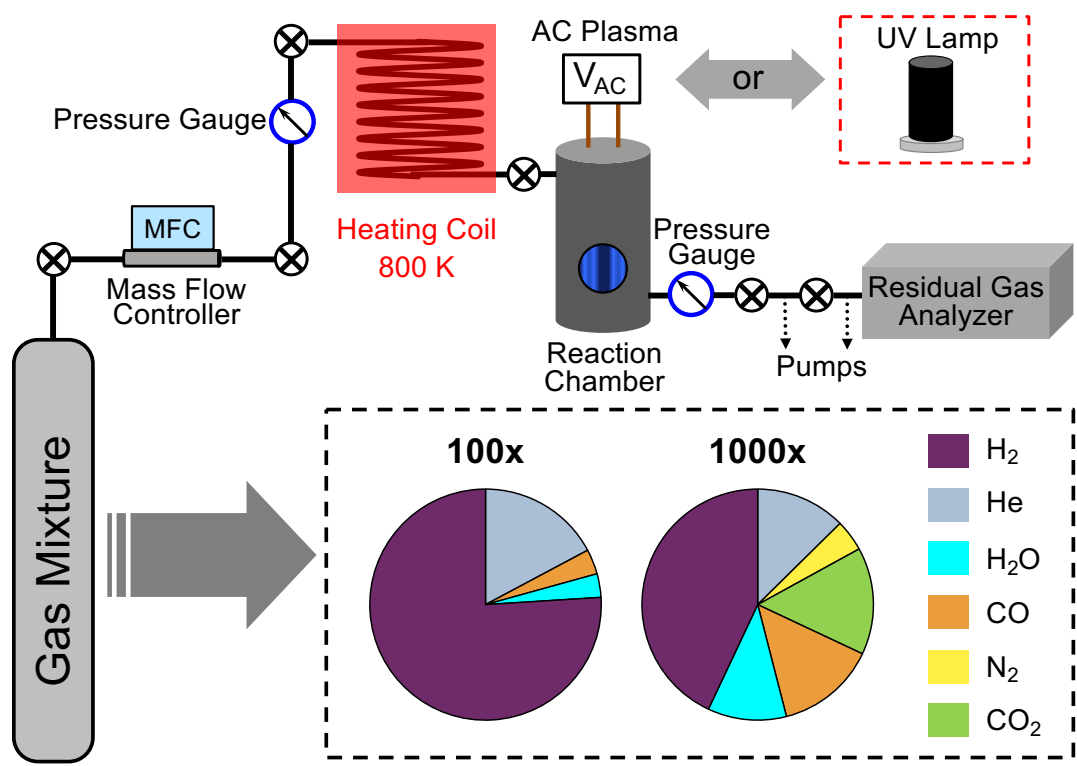

Figure 1. Simplified schematic of the PHAZER chamber and the initial gas mixtures for current work. Pie charts show the fraction of each constituent by volume mixing ratio.

## 2. Materials and Experimental Methods

*2.1. Haze Production Setup and Procedure*

We ran the experiments using the Planetary HAZE Research (PHAZER) setup (Figure 1) at Johns Hopkins University (He et al. 2017). To determine the initial gas mixtures in Figure 1, we first multiplied solar abundances for elements heavier than helium by a scale factor (100× or 1000×), commonly referred to as metallicity. Similar to our previous experiments, we then used a thermochemical equilibrium model (Moses et al. 2013a) to determine volume mixing ratios at 800 K and 1 mbar, discounting condensates and removing gas-phase species with volume mixing ratios below 1%. The mean molecular weight of our initial gas mixtures is 3.8 amu (100×) and ~15 amu (1000×), representing moderate- to high-metallicity atmospheres.

The experimental procedure is the same as in our previous studies (He et al. 2018a, 2018b, 2019, 2020; Hörst et al. 2018a) and is described here briefly. First, we prepared the initial gas mixture using high-purity gases ($H_2$-99.9999%, He-99.9995%, CO-99.99%, $N_2$-99.9997%, and $CO_2$-99.999%; Airgas) and HPLC Grade water (Fisher Chemical). We heated the well-mixed gas mixture to 800 K by flowing it through a custom heating coil and then exposed it to an energy sources (cold plasma or UV photons) for about 3 seconds in a stainless-steel chamber at a pressure of 3.2 mbar.

We used two different kinds of energy sources (cold plasma or FUV photons) to simulate different processes that occur in planetary atmospheres. Plasma generated by AC glow discharge mimics electrical activity and/or charged particles in planetary upper atmospheres, while the UV photons (110 to 400 nm) produced by the lamp simulate UV radiation from a star (Cable et al. 2012). The AC glow discharge in our experiments is created by applying voltage between two copper electrodes. The electrons are emitted from the electrodes and accelerated by the electric field. These electrons can collide with gas molecules to generate ions, metastables, and free radicals, which can further excite or ionize other gas molecules through collisions. The electrons and ions produced by the AC glow discharge are in the 5 to 15 eV range, which are energetic enough to directly dissociate $H_2$, $H_2O$, and $CO_2$, as well as triple bonded $N_2$ and CO (Cable et al. 2012). The UV lamp we used is a hydrogen light source (HHeLM-L from Resonance Ltd.), producing UV radiation with $H_2$ molecular lines in the 110 to 165 nm region (including Lyman-alpha line, 121.6 nm) and $H_2$ continuum spectrum in 165 to 400 nm. The UV output spectrum of the lamp and the absorption cross sections of the gases used in our experiments have been reported previously (Figure 3 in He et al. 2019). In this UV wavelength range (110 to 400 nm), our chamber is optically thin for both gas mixtures under our experimental conditions. The UV flux of the lamp is $\sim 3 \times 10^{15}$ photons/(sr*s). Stellar UV radiation is probably the primary source of energy for photochemistry in exoplanet atmospheres, but no existing technique in the lab can simulate the whole UV spectrum from a star including the Sun. UV lamps with similar wavelength range and flux to our lamp have been used for simulating photochemistry in planetary atmospheres (see e.g., Trainer et al. 2006, 2012; Sebree et al. 2014; Hörst et al. 2018b; Berry et al. 2019a, 2019b). UV photons in this wavelength range (110 to 400 nm) are not sufficiently energetic to directly dissociate $N_2$ or CO, but can lead to photolysis of $H_2$, $H_2O$, and $CO_2$ (He et al. 2019). The energy density of the AC glow discharge is about 170 W/m$^2$, which is ∼5 times greater than that of the UV photons, 36 W/m$^2$ (He et al. 2019).

The gas mixture flowed continuously for 72 hours at a rate of 10 standard cubic centimeters per minute. A Residual Gas Analyzer (RGA) was used to monitor the gas composition flowing out of the chamber. Solid particles were deposited on the chamber wall and on mica/quartz/glass substrates placed at the bottom of the chamber. The solid sample was collected and kept in a dry, oxygen-free $N_2$ glove box while awaiting further analysis.

*2.2. Atomic Force Microscopy (AFM) Measurements*

We examined the morphology of the deposited solid particles using a Bruker Dimension 3100 atomic force microscope. Freshly-cleaved mica substrates were used for haze particle collection because the surface of the cleaved mica is molecularly smooth. A blank mica substrate was measured with AFM to ensure that the surface of the cleaved mica is clean and smooth. The particles on mica substrates were scanned with a super sharp AFM silicon probe under ambient conditions. Details of the measurement procedure are described in He et al. (2018a, 2018b, 2020). We used tapping-mode imaging to obtain high-resolution AFM images of the particles on mica substrates (Figure 2). We measured particle sizes (diameter) with an error less than 3 nm. As in previous studies (He et al. 2018a, 2018b, 2020), we measured the particle size distribution by scanning a large area (10 μm × 10 μm) of each substrate.

*2.3. Gas Phase Composition Measurements*

We monitored gas composition during each experiment with an RGA (a quadrupole mass spectrometer). The RGA is equipped with an electron ionization (EI) source. We used a standard 70 eV ionization energy and set the scanning mass range to 1-100 amu. The background mass spectrum of the RGA chamber (at a few $10^{-7}$ mbar level) was measured before each experiment and subtracted from measured mass spectra of the gas mixtures during the experiment. Before turning on the energy source (plasma or UV), we obtained 50 mass scans of the initial gas mixture. After turning on the energy source, we waited 30 minutes before starting RGA scan to ensure that steady state of gas products in the chamber is reached. The evolving gas mixture was scanned 1000 times over the 72-hour duration of each experiment. Each scan takes ~2 minutes and the total scan time is about 34 hours. The scans recorded newly formed gas products along with remaining initial gases.

During the 1000 scans in each experiment, peak intensity at each mass varied by at most 1.0% while in steady state. To reduce noise, we averaged mass spectra of the initial gas mixture and mass spectra of the gas mixture during the experiment. Then, we normalized each average mass spectrum, using the total intensity of all peaks (1 to 100 amu) in each spectrum as a fixed reference. By comparing the normalized mass spectrum of the initial gas mixture and the gas mixture during the experiment, we identified mass peaks with significant changes (decrease or increase over 10%). Note that the measured mass spectra are not quantitative for different species due to different ionization efficiencies and instrumental responses in the mass spectrometer.

## 3. Results and Discussion

*3.1. Particles Formation and Size Distribution*

We simulated two $H_2$-rich atmospheres (100× and 1000× metallicity, Figure 1) with one of two energy sources (AC plasma or UV photons). After the experiments, we examined the substrates that were put inside the chamber during the experiments. We found that there is no visible difference between the substrates in the experiments (plasma-100×, UV-100×, plasma-1000×, and UV-1000×) and the new blank substrate, suggesting that the production rate may be relatively low. However, AFM images of the mica substrates from each experiment showed that all four experiments produced at least a layer of haze particles (Figure 2). The size and the number density of the particles differ for each case.

For both simulated atmospheres, the plasma experiments produced a higher number density of smaller particles, relative to the UV experiments. For both energy sources, particle sizes were larger for the 100× atmosphere, relative to the 1000× atmosphere. The plasma-1000× experiment produced uniformly small particles with mean diameter of 36.1 nm and the highest number density. The plasma-100× case produced larger particles (72.0 nm) with the second-highest number density. The UV experiments generate fewer particles. Particles in the UV-100× case have the largest mean diameter (82.8 nm) and spread over the surface, while particles in the UV-1000× case tend to aggregate. AFM images show that the particle monomers are in spherical shape for all the cases. In the UV-1000× case, many of the spherical monomers assemble to irregular aggregates with projected-area-equivalent diameter from 78 nm to 436 nm (projected-area-equivalent diameter corresponding to the diameter of a sphere with the same projected area as the particle, Pabst & Gregorova 2007).

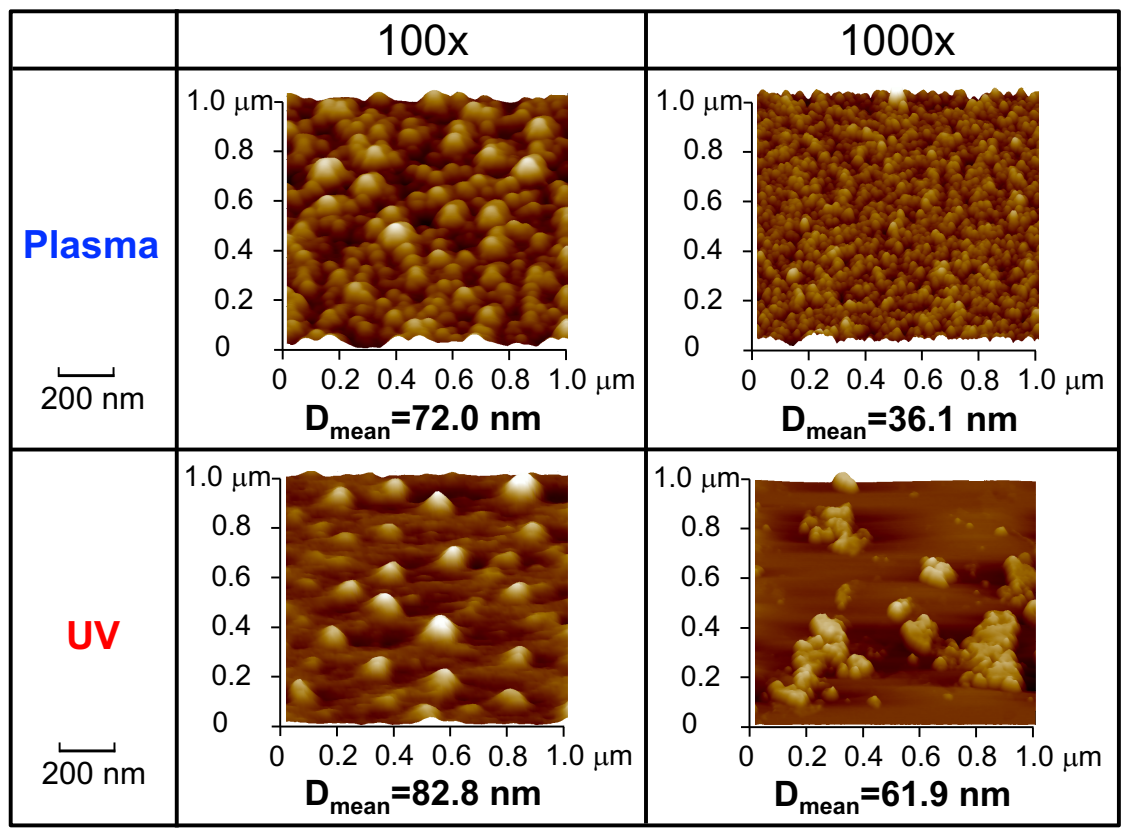

Figure 2. AFM images of particles on the mica substrates in a 1 μm × 1 μm scanning area for each experiment. The mean diameter ($D_{mean}$, nm) of the haze particles (monomers) is shown under each image.

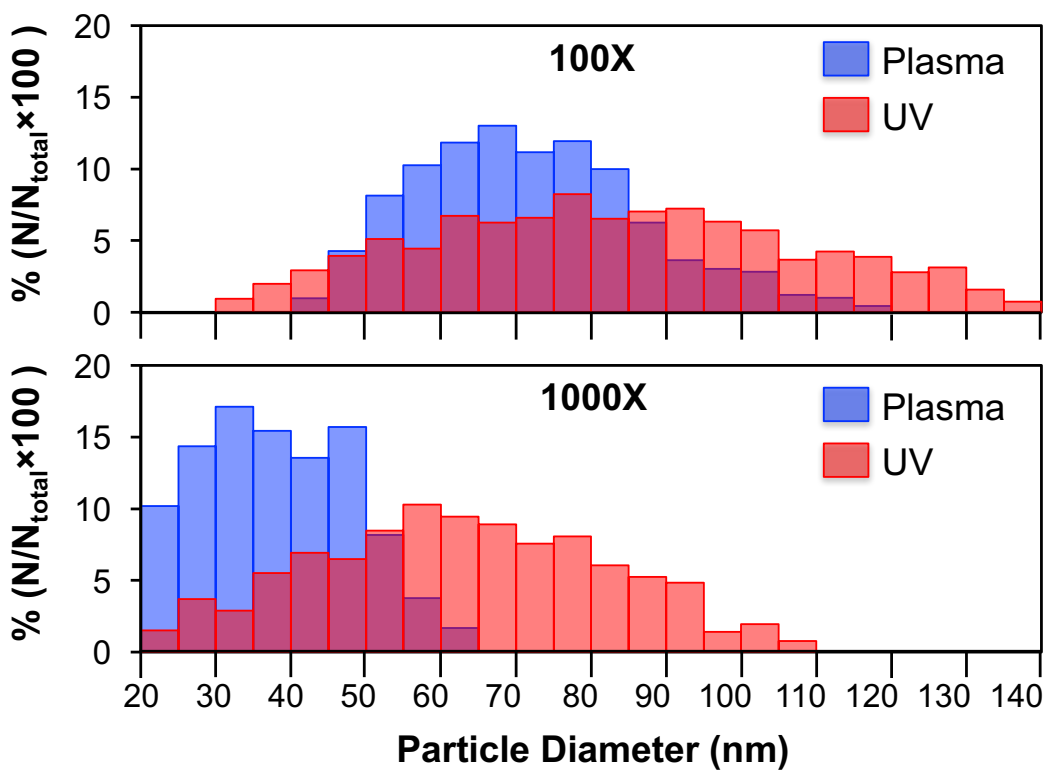

Figure 3. Size distribution of the haze particles formed in the plasma (blue) and UV (red) experiments, for 100× metallicity (top), or 1000× metallicity (bottom). The number of particles used for the size-distribution statistics are 34787 (plasma-100×), 7520 (UV-100×), 288270 (plasma-1000×), and 17754 (UV-1000×). The UV experiments produced a broader range of particle sizes.

Figure 3 shows the particle size distribution for each film, measured over a larger area (10 μm × 10 μm). The 100× atmosphere yields larger particle sizes than the 1000× atmosphere, particularly for the plasma source. The UV experiments produced wider size distributions than the plasma experiments, which is consistent with previous studies (He et al. 2018a, 2020). Both the energy density and energy type could affect the size range of the particles. With plasma as energy source, higher energy density produces more nuclei center and supplies sufficient new species for more uniform particle growth. The lower energy density of the UV photons creates fewer new species, leading to fewer nucleation centers and slower growth rate; localized growing around each nucleation center could cause nonuniform growth to generate different size of particles. Gas-solid heterogeneous reactions are more likely to happen in the UV experiments and could also play a role in the formation of a wider size range of particles. Beyond the energy density, different types of energy sources initiate different reactions, resulting in haze particles with different chemical compositions. Different chemical structures could lead to different morphologies of the particles. In addition, the physical properties of different compounds, such as vapor pressure, polarity, viscosity, cohesion, adhesion, and surface energy, could also affect the growth of the particles. However, the nucleation and growth of particles is a complex process, and studying the mechanism is beyond the scope of this paper. The general size range (20 to 140 nm) of the four experiments here is similar to our previous experiments for different temperatures and metallicities (He et al. 2018a, 2018b, 2020). Such small particles will scatter short wavelengths efficiently, affecting the observed spectra of an exoplanet. Such small particles will scatter short wavelengths efficiently, affecting the observed spectra of an exoplanet. The size range lies in the Rayleigh scattering regime for visible and infrared (IR) photons. These small particles can lead to a Rayleigh scattering slope of observed transmission spectra and weaken spectral features in short wavelengths (Moran et al. 2018). The result from the UV-1000× case suggests that these small particles could grow to larger aggregates that lie in Mie scattering regime (particle diameters approximately equal to the wavelength of the incident light). Besides scattering, these haze particles also absorb light in different wavelengths and could have distinguishable absorption features in the observed spectra. Further investigation of their optical and compositional properties is required to understand their impacts on the observation.

*3.2. Haze Production Rate*

Due to relatively low production rates, we were unable to collect macroscopic amounts of solid samples. In order to compare the particle production efficiency, we estimated the production rate on the basis of the size distribution (He et al. 2018a, 2020). We first calculated the total volume ($V$) of all the particles produced in each experiment by assuming that the particles are deposited on the wall of the chamber in a uniform manner:

$$V = \sum_{i=0}^{i} \frac{\pi}{6} D_i^3 N_i \qquad (1)$$

where $D_i$ is the median particle diameter in each bin and $N_i$ is the number of particles in each bin over the total available surface area within the chamber. With the total volume and an assumed particle density ($\rho$ = 1.38 g cm$^{-3}$, same as a Titan tholin sample, He et al. 2017), we can estimate the total mass (mg) and the production rate (mg h$^{-1}$, Figure 4) of each experiment. It is possible that there are multiple layers of particles deposited on the wall of the chamber, but we only consider one uniform layer. Therefore, the total volume/mass or the production rate we calculated gives a lower limit.

In these 800 K experiments, the particle production rates with plasma are about 3 times higher than those with UV for both simulated atmospheres (Figure 4). The production rates per volume are calculated by dividing the value shown in Figure 4 (mg h$^{-1}$) by the volume of the reaction chamber (2160 cm$^3$), which are 7.4 × 10$^{-5}$ mg cm$^{-3}$ h$^{-1}$ for the plasma-100× case, 2.4 × 10$^{-5}$ mg cm$^{-3}$ h$^{-1}$ for the UV-100× case, 9.7 × 10$^{-5}$ mg cm$^{-3}$ h$^{-1}$ for the plasma-1000× case, and 3.1 × 10$^{-5}$ mg cm$^{-3}$ h$^{-1}$ for the UV-1000× case. The higher production rate with the plasma energy source is consistent with our previous studies (He et al. 2018a, 2020), reflecting the higher energy density of the plasma source relative to the UV source. The energy density of the plasma in our experiments is estimated to be ~5 times higher than that of the UV radiation (He et al. 2019). The production rate is positively related to the energy density, but it is not a linear relationship because the plasma and UV are two different types of energy sources, which can initiate different chemistry in the gas mixture. For instance, the UV photons in our experiments are not energetic enough to directly break strong bonds, such as those in $N_2$ or CO, which may also affect the production rate and composition of the haze particles. It requires further investigations to quantify the dependence of the energy density on the production rate, such as experiments using UV photons with the same wavelength range but different flux levels.

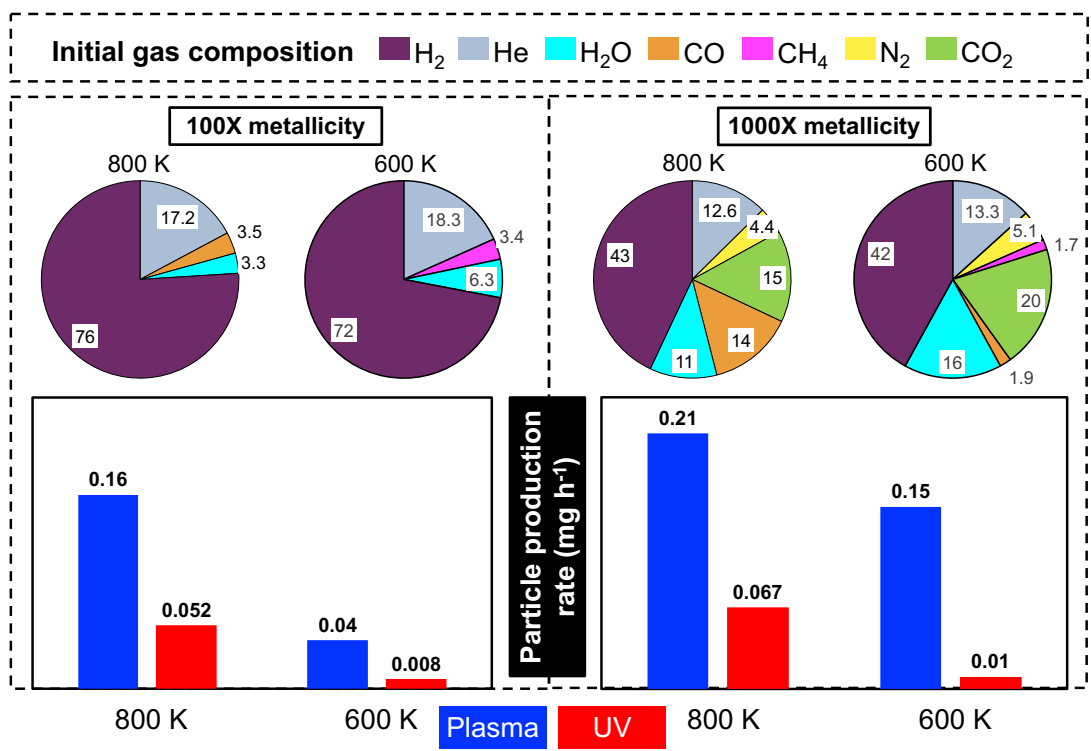

Figure 4. The initial gas mixtures and the haze particle production rate (mg h$^{-1}$) in the 800 K experiments with plasma and UV energy sources. The plasma experiments have higher haze production rates than the UV experiments for both simulated atmospheres (100× or 1000× metallicity). The uncertainties (due to the measurement error of the particle diameter and the uncertainty of the particle number counting) of the estimated production rates are 3.4% (plasma-100×), 2.3% (UV-100×), 4.2% (plasma-1000×), and 3.5% (UV-1000×), respectively. Those from previous 600 K experiments are shown for comparison.

The production rate of the 1000× metallicity case is higher than that of the 100× metallicity case with both energy sources, due to the compositional difference between the two simulated atmospheres. As shown in Figure 4, the 1000× metallicity case contains higher concentration of CO (14% versus 3.5%), and two extra gases ($CO_2$ and $N_2$). $CO_2$ and extra CO can provide more carbon sources for producing organic hazes, while nitrogen from $N_2$ could be also incorporated into the solid haze particles (Imanaka & Smith 2010, He & Smith 2013, 2014; Carrasco et al. 2015; Sebree et al. 2016; Hicks et al. 2016; Gautier et al. 2017; Hörst et al. 2018b). There is less $H_2$ (43% versus 76%) in the 1000× metallicity case, but the fraction is still more than sufficient to provide a reducing environment in which organic molecules can form. Photochemical models (Zahnel et al. 2016) suggest that sulfur photochemistry could be important in similar conditions (700 K, 1000× metallicity), and laboratory studies (He et al. 2020, Reed et al. 2020) show that small amount of $H_2S$ could enhance haze production rate significantly. In the near future, we will include $H_2S$ in the initial gas mixture and investigate sulfur chemistry in $H_2$-rich atmospheres, as we did previously for $CO_2$-rich atmospheres (He et al. 2020).

The two $H_2$-rich atmospheres simulated here are similar to the $H_2$-rich atmospheres in our previous 600 K experiments (He et al. 2018a, Hörst et al. 2018a), shown in Figure 4. For the 100× metallicity case, $H_2$, He, and $H_2O$ account for over 96% of the gas mixture in both 600 K and 800 K experiments, but the 800 K experiment here contains CO as carbon source instead of $CH_4$. For the 1000× metallicity case, the 600 K experiment includes extra $CH_4$ (1.7%). However, the production rates here in the 800 K experiments are higher than the 600 K experiments for both the 100× and 1000× metallicity cases with either of two energy sources (plasma and UV). The higher production rate without $CH_4$ indicates that $CH_4$ is neither required to generate organic hazes nor necessary to promote the organic haze formation. CO and/or $CO_2$ can serve as a carbon source for haze formation (Fleury et al. 2017, 2019; He et al. 2018a, 2018b, 2019, 2020; Hörst et al. 2018b; Moran et al. 2020). Our study was intended to simulate photochemical haze formation in warm $H_2$-rich atmospheres of sub-Neptunes around M dwarfs, since this type of exoplanets are more common in our galaxy and will be the main targets for the near future observations. However, similar processes could happen in any warm $H_2$-rich atmospheres as long as the atmospheric environment is sufficient. Therefore, our results can inform haze formation processes in atmospheres of other type of exoplanets such as warm Jupiters.

Molecules at higher temperature (800 K versus 600 K) are more reactive because higher thermal velocities imply more frequent collisions between molecules, thereby increasing the reaction rates and production rates of haze particles. However, at higher temperature, certain species would remain as gases rather than condensing out as solid particles. Common organic molecules are not stable at very high temperature (>1000 K, Johns et al. 1962, Smay 1985), so we would not necessarily expect significant organic haze in the atmospheres of Jupiters hotter than 1000 K. This is consistent with a recent modeling study (Gao et al. 2020), which found that aerosol composition is dominated by silicates for hot giant exoplanets with planetary equilibrium temperatures above 950 K, while is dominated by hydrocarbon aerosols below 950 K. However, it is possible to form organic hazes at temperatures above 1000 K as reported by Fleury et al. (2019) who found that photochemistry can lead to the formation of an organic solid condensate at 1500K. We also need to consider that temperatures on the night sides of these hot planets can be significantly lower than on the day side, so there may be situations where gases that are photochemically produced on the day side are transported by winds to the night side, where they can form condensates.

*3.3. Mass Spectra of Gas Phase Products*

Figure 5 shows recorded mass spectra of the gas phase before and while the energy source (plasma or UV) was on. We identified mass peaks that changed significantly (>10%). In general, peaks that decrease are associated with initial gas molecules that are dissociated due to the discharge or UV photons, while peaks that increase indicate newly formed gas products. Note that some of peaks associated with initial gas molecules decreases less than 10% due to their high mixing ratio in the initial gas mixture, but their intensity changes are significant. Table 1 lists the changed molecular peaks and the relative intensity changes along with the associated species. Their percentage differences from the original gas mixture are also included in Table 1.

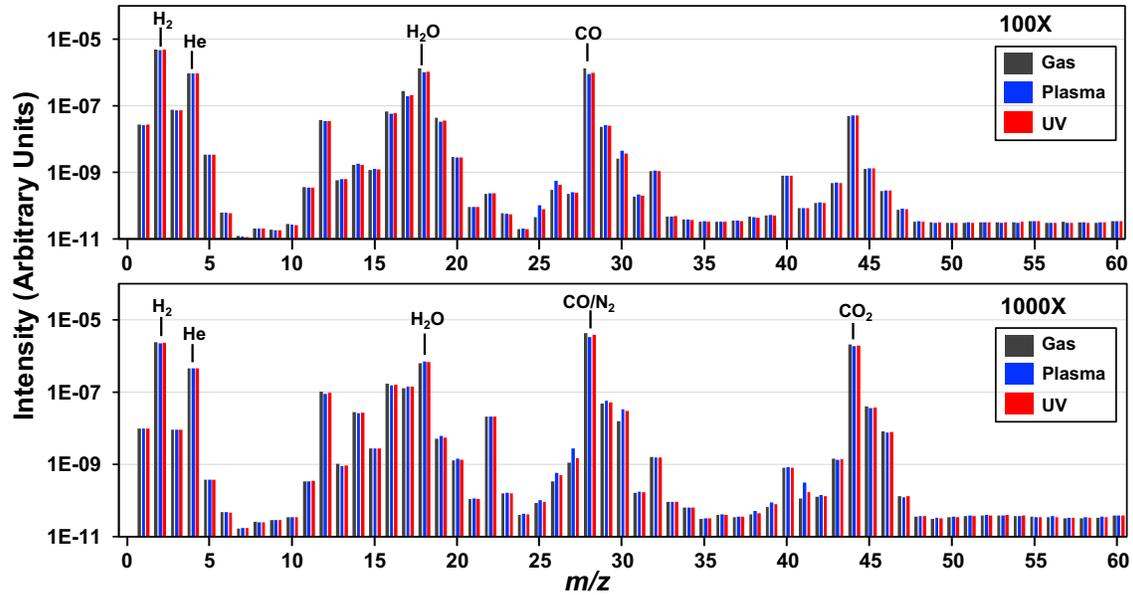

Figure 5. The mass spectra of the gas mixtures in the 100× (top) and 1000× (bottom) metallicity experiments: initial gas mixture (gray), plasma on (blue), and UV on (red). The initial gas compositions are labeled in the spectra. The decreased peaks indicate the initial gas molecules dissociated under energy exposure while the increased ones are due to the newly formed gas products.

For the 100× metallicity case, the intensities of $H_2$, $H_2O$, and CO in the initial gas mixture decrease in both the plasma and UV experiments. The increased peak at 26 amu is acetylene ($C_2H_2$), while the peak at 30 amu could be from ethane ($C_2H_6$) and/or formaldehyde (HCHO). For the 1000× metallicity case, $H_2$, CO, $N_2$, and $CO_2$ all decrease under energy exposure. The reactions of $CO/CO_2$ with $H_2$ could produce additional $H_2O$ molecules in the system, increasing the intensity of $H_2O$. Besides $C_2H_2$ and $C_2H_6$/HCHO, several nitrogen-containing molecules are formed in the gas phase, such as hydrogen cyanide (HCN) at 27 amu, acetonitrile ($CH_3CN$) at 41 amu, and probably methanimine ($CH_2NH$) at 29 amu and nitric oxide (NO) at 30 amu. Our RGA is a unit-resolution quadrupole mass spectrometer, so it cannot resolve species with identical nominal mass. Therefore, CO and $N_2$ both contribute to the peak at 28 amu for the 1000× metallicity case; the peak at 30 amu could have contributions from $C_2H_6$/HCHO for the 100× metallicity case and from $C_2H_6$/HCHO/NO for the 1000× metallicity case.

**Table 1. Assignments and the relative intensity change of the changed molecular peaks in the mass spectra. The uncertainties (due to the measurement error of the**

**mass spectra) for the intensity changes are less than 1%. The values in the parentheses are the percentage changes of these peaks compared to the initial gas mixtures.**

| 100× metallicity | | | |
|---|---|---|---|
| Peaks ($m/z$) | Species | Plasma | UV |
| 2 | $H_2$ | ↓ $1.5×10^{-7}$ (3.2%) | ↓ $1.1×10^{-7}$ (2.3%) |
| 18 | $H_2O$ | ↓ $3.0×10^{-7}$ (22.8%) | ↓ $2.6×10^{-7}$ (20.1%) |
| 28 | CO | ↓ $4.3×10^{-7}$ (31.8%) | ↓ $3.6×10^{-7}$ (27.4%) |
| 26 | $C_2H_2$ | ↑ $2.6×10^{-10}$ (86.1%) | ↑ $1.2×10^{-10}$ (39.9%) |
| 30 | $C_2H_6$/HCHO | ↑ $2.0×10^{-9}$ (74.1%) | ↑ $1.1×10^{-9}$ (41.7%) |
| Total | | ↑ $2.3×10^{-9}$ | ↑ $1.2×10^{-9}$ |
| 1000× metallicity | | | |
| Peaks ($m/z$) | Species | Plasma | UV |
| 2 | $H_2$ | ↓ $1.9×10^{-7}$ (7.8%) | ↓ $1.2×10^{-8}$ (5.2%) |
| 28 | CO/$N_2$ | ↓ $8.7×10^{-7}$ (20.5%) | ↓ $4.2×10^{-7}$ (12.5%) |
| 44 | $CO_2$ | ↓ $2.3×10^{-7}$ (10.9%) | ↓ $1.4×10^{-7}$ (7.7%) |
| 26 | $C_2H_2$ | ↑ $2.4×10^{-10}$ (70.3%) | ↑ $1.6×10^{-10}$ (28.1%) |
| 27 | HCN | ↑ $1.7×10^{-9}$ (147.6%) | ↑ $3.5×10^{-10}$ (12.7%) |
| 29 | $CH_2NH$ | ↑ $9.6×10^{-9}$ (19.7%) | ↑ $2.7×10^{-9}$ (4.75%) |
| 30 | $C_2H_6$/HCHO/NO | ↑ $1.8×10^{-8}$ (114.6%) | ↑ $1.5×10^{-8}$ (44.6%) |
| 41 | $CH_3CN$ | ↑ $2.0×10^{-10}$ (176.2%) | ↑ $5.6×10^{-11}$ (17.4%) |
| Total | | ↑ $3.0×10^{-8}$ | ↑ $1.8×10^{-8}$ |

↓: Decrease; ↑: Increase.

For a given initial gas mixture (100× or 1000× metallicity case), the two different energy sources produce the same types of new gas products, but the quantities of these new species are higher in the plasma experiment than in the UV experiment. In the plasma experiments, the higher yield of the gas products is consistent with the higher production rate of solid particles. The estimated energy density of the plasma energy source (170 W/m$^2$) is about 5 times higher than that of the UV lamp (36 W/m$^2$), as reported previously (He et al. 2019, 2020). Both energy sources have higher energy density than a hypothetical warm (800 K) exoplanet (~14 W/m$^2$ in the range of 1 to 300 nm, which is important for atmospheric photochemistry) around a given host M-star (3000 K) (He et al. 2020). The plasma source is able to break strong bonds in CO and $N_2$, while the UV photons generated by our UV lamp cannot. However, we detected nitrogen-containing organic molecules with either energy source when $N_2$ is included in the gas mixture. The nitrogen incorporation into organic molecules (in both the gas phase and solid phase) with similar far UV lamp have been observed in previous studies (Hodyss et al., 2011; Trainer et al., 2012; Yoon et al., 2014; Hörst et al. 2018b; Berry et al. 2019a, 2019b). Several possible photochemical processes for nitrogen incorporation have been suggested (Trainer et al. 2012; Yoon et al. 2014; Berry et al. 2019a), but the exact mechanism is still unclear.

Both simulated atmospheres, 100× metallicity and 1000× metallicity, are $H_2$-rich atmospheres, but the 1000× metallicity case has more CO (14% versus 3.5%) and also has additional $CO_2$ and $N_2$. The total amount of the new gas products in the 1000× metallicity case is more than 10 times higher than that of the 100× metallicity case with both energy sources. Note that we were only able to detect the most abundant products in the gas phase, which are the small molecules (less than three carbons) listed in Table 1. Most of the newly-formed gas products are photochemically active and tend to further react to form large molecules. Medium-sized and large molecules are thus expected to be formed in the gas phase, but their abundances might be lower than the detection limit of the RGA. These new gas products could serve as key precursors for producing more complex compounds and haze particles, and further compositional analysis of the haze particles is needed to understand possible haze formation mechanisms. More gas products lead to the higher production rate of solid particles in the 1000× metallicity case. In addition, with $N_2$ in the initial gas mixture, several new gas products are nitrogen-containing species. The new gas products are indicative of the composition of the solid haze particles; the formation of nitrogen-containing molecules in the gas phase suggests that nitrogen may also be incorporated into solid particles (Imanaka & Smith 2010, He & Smith 2013, 2014; Carrasco et al. 2015; Sebree et al. 2016; Hicks et al. 2016; Gautier et al. 2017; Hörst et al. 2018b; Berry et al. 2019a, 2019b; Ugelow et al. 2020). Gas molecules will be easier to detect than complex haze compositions with remote sensing techniques. JWST and future ground-based facilities will be able to probe spectral features of major gas compositions. These gas molecules may serve as important atmospheric chemical indictors of photochemistry and haze formation in exoplanet atmospheres.

The carbon-to-oxygen ratio (C/O) is used as an important proxy to classify exoplanet atmospheres, as carbon and oxygen are among the most abundant elements in the Universe and the C/O ratio could be related to the bulk atmospheric chemical composition (Madhusudhan 2012). Many modeling studies have investigated the effect of the C/O ratio on the atmospheric properties and observable spectra of transiting exoplanets (e.g., Madhusudhan et al. 2011, Moses et al. 2013b). Its impact on photochemistry and hydrocarbon haze formation in warm (<1000 K) exoplanet atmospheres was recently reported (Kawashima & Ikoma 2019). Here our two simulated $H_2$-rich atmospheres have similar C/O ratio (~0.5). However, the cases lead to different results in terms of haze particles size, haze production rate, and gas product yield. Our results show that distinct chemical processes occur in atmospheres with similar C/O ratio, which could influence the observable spectra of exoplanet atmospheres.

Our two simulated atmospheres have quite different C/H (0.02 versus 0.45) and N/C (0 versus 0.3) ratios, which may be responsible for distinct chemical processes happening in the atmospheres. The full set of elemental abundances in an atmosphere should be considered, as suggested in Drummond et al. (2019). The elemental abundances are important, but the chemical bonding between the elements could also significantly affect the chemical reactivity of molecules and how they interact with light. For instance, $CH_4$, CO, and $CO_2$ can all provide carbon for producing organic molecules, but their efficiency is not equal ($CH_4$ > CO > $CO_2$, based on their bonding environments (Darwent 1970, Blanksby & Ellison 2003)). Likewise, for a given energy source, nitrogen in $NH_3$ is easier to incorporate into organic molecules than nitrogen in $N_2$. Therefore, elemental abundances and bonding environments are both important when modelling or interpreting the observations of exoplanet atmospheres.

The two simulated atmospheres (100× and 1000× metallicity) do not contain $CH_4$ in the initial gas mixtures. However, photolysis of $CH_4$ is often considered as the main pathway to generate organic haze in photochemical models (e.g. Liang et al. 2004, Zahnle et al.

2009, Moses et al. 2011; Miller-Ricci Kempton et al. 2012), and exoplanet studies usually estimate the production rate of organic haze only from $CH_4$ photodissociation (e.g. Morley et al. 2013, Gao et al. 2020). The production rates in the 800 K experiments (without $CH_4$) are higher than the 600 K experiments (with $CH_4$), demonstrating that $CH_4$ is not necessary for organic haze formation. Alternative photochemical pathways starting with CO and/or $CO_2$ could be important for organic haze formation, and should also be considered in models.

## 4. Conclusions

We investigated haze formation in warm (800 K) $H_2$-rich exoplanet atmosphere analogs (100× and 1000× metallicity) with energy input from AC plasma or a UV lamp. We find that small haze particles (20 to 140 nm) are produced in both simulated atmospheres with either energy source. The haze production rate ($2.4 \times 10^{-5}$ to $9.7 \times 10^{-5}$ mg cm$^{-3}$ h$^{-1}$) is relatively low compared to our standard Titan experiments ($3.4 \times 10^{-3}$ mg cm$^{-3}$ h$^{-1}$). The mass spectra reveal that there are more kinds of new gas products formed in the 1000× metallicity case and their total relative yield is also more than 10 times higher in the 1000× metallicity case than in the 100× metallicity case. Our results show that the gas phase chemistry, solid particle size, and production rate are all dependent on the initial gas mixtures. CO and $N_2$ in the initial gas mixture can induce complex chemical reactions in $H_2$-rich atmospheres and form reactive gas products, which can serve as key precursors for producing more complex compounds and solid haze particles. This study demonstrates that both the elemental abundances and the bonding environments of an atmosphere could play important roles in photochemistry of exoplanet atmospheres, which should be taken into account when modelling or interpreting the observations of exoplanet atmospheres. More laboratory work is necessary to investigate the role of each molecule in different background atmospheres.

## Acknowledgements

This work was supported by the NASA Astrophysics Research and Analysis Program NNX17AI87G. X. Yu is supported by a 51 Pegasi b Fellowship.